\documentclass[twocolumn,showpacs,preprintnumbers,amsmath,amssymb,pre]{revtex4}
\usepackage{times}
\usepackage[sort&compress]{natbib}
\usepackage[dvips]{graphicx}

\usepackage{color}

\begin{document}

\affiliation{$^2$}


\title{Biological Evolution in a Multidimensional Fitness Landscape}
\author{David B. Saakian$^{1,2,3}$}
\email{saakian@phys.sinica.edu.tw}
\author{Zara  Kirakosyan$^{2}$}
\author{Chin-Kun Hu$^{1}$}
\email{huck@phys.sinica.edu.tw}

\affiliation{$^1$Institute of Physics, Academia Sinica, Nankang,
Taipei 11529, Taiwan}

 \affiliation{$^2$Yerevan Physics
Institute, Alikhanian Brothers St. 2, Yerevan 375036, Armenia}

\affiliation{$^3$National Center for Theoretical Sciences (North):
Physics Division, National Taiwan University, Taipei 10617,
Taiwan}


\date{\today}

\begin{abstract}
We considered a {multi-block} molecular model of biological
evolution, in which fitness is a function of the mean types of
alleles located at different parts (blocks) of the genome. We
formulated an infinite population model with selection and mutation,
and calculated the mean fitness. For the case of recombination, we
formulated a model with a multidimensional fitness landscape {the
dimension of the space is equal to the number of blocks}) and
derived a theorem about the dynamics of initially narrow
distribution. We also considered the case of lethal mutations.
We also formulated the finite population version of the model in the
case of lethal mutations.{ Our models, derived for the virus
evolution, are interesting also for the statistical mechanics and
the Hamilton-Jacobi equation as well.}
\end{abstract}



\maketitle


\section{Introduction}

The investigation of biological evolution models
\cite{ei71,ss82,ei89,ck70,pe97}  is one of the most fruitful
applications of statistical mechanics or theoretical physics to
biological problems \cite{bg00,dr01}-\cite{de07}. To solve the
evolution models, one can apply the whole machinery of modern
theoretical physics: spin-glass physics methods \cite{fp97}, quantum
statistical mechanics \cite{ba97,sa04,sa04b,sa06,sa06a}, quantum
field theory \cite{sa04,sa06a}, Hamilton-Jacobi equation (optimal
control) \cite{sa07,07preKaneko,08pre06}.


The genome, a collection of genes with different types, could be
considered as a particular spin configuration of a statistical
system, where the fitness (the rate to produce offsprings of the
given genome) is equivalent to the Hamiltonian of the spin system.
In evolution theory, the notion of fitness
 is central
in defining the general features of evolution or in modeling a
concrete experiment.  Fitness is a complicated function of gene
content (types of genes) of the genome in sequence space; this
function is assumed to have a mean-field like behavior. Most of the
investigations have been devoted to the symmetric fitness case, when
there is a master (reference) sequence, and fitness (energy) is a
simple function of the Hamming distance from that sequence
\cite{ss82}. In \cite{sa06a},  a generalization of symmetric fitness
landscape was considered, when there are some $K$ reference
sequences, and the fitness was a function of $K$ Hamming distances
from these reference sequences.  In \cite{pe96,or06,ja10}, there
were suggested evolution models where the genome consisted of
different blocks and the fitness is a function of the gene mean
types at different blocks. In the current article, we follow the
idea of \cite{pe96}, considering an infinitely long genome, a
collection of a finite number of blocks, defining mean
"magnetizations" at any such block and the fitness as a function of
block magnetizations. We then use the Hamilton-Jacobi equation
\cite{sa07} to solve the equation. This new approach is technically
easier than that used in \cite{sa06a}. Thus in the present paper, we
can calculate the mean fitness of a recombination model in a
multidimensional fitness landscape.

Recombination is one of the key factors in evolution. The
mathematical aspects of recombination were analyzed in
\cite{ew04,bu00,ba03}. Recently, there was good
 progress in the statics of recombination
\cite{le05,de07} and there was some advance in the dynamics
\cite{sa10b}. We will formulate the recombination model in a
multidimensional fitness landscape for many-loci haploid model with
two alleles (type of gene) at any locus (position of a gene in the
genome).

 The rest of the paper is organized as follows:  In section II, we formulate
 and solve (calculate the mean fitness) the evolution model with selection
 and mutation in a multi-dimensional fitness landscape, including the
 case of lethal mutations \cite{sa10c,11PlosOne}. We consider 2 block models
 for the lethal mutations and an asymmetric initial distribution.
  In section III, we formulate the
 recombination model in a multidimensional space. While we could
 not calculate the mean fitness, we derive a general result regarding the
 dynamics of population for the initial narrow distribution.
 In Sec. IV, we summarize our results and
 discuss problems for further research.

\section{
The multidimensional model}
\subsection{The Model}

We identify the alleles as spins and consider the genome as a
collection of $L$ spins taking the values $\pm 1$. In the peak
configuration, all spins take value +1. Our model is a simple
generalization of the Crow-Kimura model \cite{ck70,ba97}.
 The genome is
a collection of $H$ pieces (blocks), with the length $L_{n},1\le
n\le H$, such that $\sum_{n=1}^{H} L_{n}=L$.

 Any sequence is characterized
by $l_1, \dots, L_H$, the number of "-" (negative) spins in the
blocks. We introduce  the "magnetization" $m_n$, defined as
\begin{eqnarray}
 \label{e1} m_n= 1-\frac{2l_{n}}{L_{n}},
\end{eqnarray} at
the $n$-th piece of genome for all of $n$ with $1 \le n\le H$. Our
fitness $r$ is a function of $(l_1,\dots,l_H)$. Thus, we define
$r_{l_1,\dots,l_H}\equiv Lf(m_1,\dots,m_H)$. {The discrete
variables $l_n$ are defined in the interval  $[0,L_n]$, while the
magnetizations $m_n$ are becoming continuous at the limit $N\to
\infty$ and $-1\le m_n\le 1$. } We define the function
$f(m_1,\dots,m_H)$ as a fitness function.

The description of the mutation process is the principal point in
the  definition of the model. In order to describe mutations we use
the coefficients  $ x_{n\pm}(l_n,L_n)$:
\begin{eqnarray} \label{e2}
 x_{n+}(l_n,L_n)=\frac{l_n}{L_n},~~  x_{n-}(l_n,L_n)=\frac{L_n-l_n}{L_n},
\end{eqnarray}
where  $L_n$   denotes the length of the $n$-th  piece,
$x_{n+}(l_n,L_n)$ and $x_{n-}(l_n,L_n)$ are { the fractions of $-$
and $+$ spins in the  $n$-th  piece.}

  If the initial
distribution of the population is symmetric, i.e. all the sequences
with the same $l_n$ having the same probability,{ we describe the
system through the $p(l_1,\dots,l_H,t)$, the probability of all
sequences having $l_1, \dots,l_H$ minus spins in corresponding
blocks}. Then we write the following system of equations:
\begin{eqnarray}
\label{e3} &&\frac{dp(l_1,\dots,l_H,t)}{dt}=
(r_{l_1,\dots,l_H}-LR)p(l_1,\dots,l_H,t)\nonumber\\
&& -p(l_1,\dots,l_H,t)
 \nonumber\\
&&+\sum_{\beta=\pm 1,~n}
L_nx_{\beta}(l_n,L_n)p(l_1,\dots,l_n-\beta,\dots,l_H,t),\nonumber\\
&&   R=\frac{1}{L}\sum_{0\le l_n\le
L_n}p(l_1,\dots,l_H,t)r_{l_1,\dots,l_H},
\end{eqnarray}
The sum over $n$ extends from 1 to $H$ and $R$ and is  the mean
fitness. We considered  mutations independently at all pieces of the
genome; thus, in the middle line at the right hand side of Eq.
(\ref{e3}) we changed an $l_n$ at the position $n$, where $n$
changes from 1 (first piece of genome) until $H$ (the last piece of
genome).

The current configuration $(l_1,\dots,l_n,\dots,l_H)$ could be
obtained from either  $(l_1,\dots,l_n+1,\dots,l_H)$, reversing one
of $l_n+1$ $"-"$ spins, or from the $(l_1,\dots,l_n-1,\dots,l_H)$
configuration reversing one of $(L_n-l_n+1)$ $"+"$ spins. There
are $(l_n-1)$ such possibilities for the first case, and
$(L_n-l_n+1)$ for the second case. Dividing by $L$, we derived the
coefficients $x_{-}(l_n+1,L_n)$ and $x_{+}(l_n-1,L_n)$ in
Eq.(\ref{e3}). For $H=1$, Eq.(\ref{e3}) coincides with the
Crow-Kimura model \cite{ck70,bw01,sa04b}.

Let us consider the linear part of the latter equation, and write
an equation for $P(m_1,\dots,m_H,t)\equiv p(l_1,\dots,l_H,t)$
\begin{eqnarray}
\label{e4} &&\frac{dP(m_1, \dots, m_H,t)}{dt}\nonumber\\
&&= L(f(m_1,\dots,m_H)-1)P(m_1,\dots,m_H,t)
 \nonumber\\
&&+\sum_{\beta=\pm 1,1\le n\le H}
    L_n(\frac{1+\beta m_n}{2}+\frac{\beta-1}{L_n})\nonumber\\
 && \times   P(m_1,\dots,m_n+\frac{2\beta}{L_n},\dots,m_H,t).
\end{eqnarray}
Following \cite{tb74,jer75}, we define the mean fitness in the
steady state of Eq.(\ref{e3}) as the largest eigenvalue of the
quadratic matrix on the left hand side of Eq.(\ref{e4}).

Following \cite{sa07}, we assume an anzats:
\begin{eqnarray}
\label{e5} P(m_1,\dots,m_H,t)=\exp[Lu(m_1,\dots,m_H,t)].
\end{eqnarray}
Then with $1/L$ accuracy we get the following Hamilton-Jacobi
equation (HJE):
\begin{eqnarray}
\label{e6} &&\frac{\partial u(m_1,\dots,m_H)}{\partial t}=-
H(m_1,\dots,m_H; \frac{\partial
u}{\partial_{m_1}},\dots,\frac{\partial u}{\partial_{m_H}}),
\nonumber\\
&& -H(m_1,\dots,m_H;\hat P_1,\dots,\hat P_H)=f(m_1,\dots,m_H)
 \nonumber\\
&&-1+\sum_{1\le n\le H} \frac{L_n}{L}(\frac{1+m_n}{2}e^{2\hat
P_n}+
   \frac{1-m_n}{2}e^{-2\hat P_n}),
\end{eqnarray}
where we  missed  $O(1/L)$ terms and introduced the momentums
$\hat P_n=\partial u/\partial m_n$.

Consider the asymptotic solution:
\begin{eqnarray}
\label{e7} u(m_1,\dots,m_H,t)=Rt+ u_0(m_1,\dots,m_H),
\end{eqnarray}
we get an equation
\begin{eqnarray}
\label{e8} R&=&f(m_1,\dots,m_H)-1\nonumber\\ &+&\sum_{1\le n\le H}
[\frac{L_n}{L}\frac{1+m_n}{2}e^{2\frac{\partial
u_0(m_1,\dots,m_n,\dots,m_H)}{\partial_{m_n}}}\nonumber\\
   &&+\frac{L_n}{L}\frac{1-m_n}{2}e^{-2\frac{\partial
   u_0(m_1,\dots,m_n,\dots,m_H)}{\partial_{m_n}}}].
\end{eqnarray}
On the other hand, we have a condition that at any point $m$, our
$R$ should be higher than the minimum of the right hand side,
considered as a function of momentums $\frac{\partial
u_0(m_1,\dots,m_n,\dots,m_H)}{\partial_{m_n}}$. We define
$$
U(m_1,\dots,m_H)=min[f(m_1,\dots,m_H)-1+$$
$$\sum_{1\le n\le H}
[\frac{L_n}{L}\frac{1+m_n}{2}e^{2\frac{\partial
u_0(m_1,\dots,m_n,\dots,m_H)}{\partial_{m_n}}}$$
$$+\frac{L_n}{L}\frac{1-m_n}{2}e^{-2\frac{\partial
   u_0(m_1,\dots,m_n,\dots,m_H)}{\partial_{m_n}}}]$$
Examining the solution of the minimum problem and looking at
different points $m$, we find:
\begin{eqnarray}
\label{e9}&&R\ge max[U(m_1,\dots,m_n)]|_{m_1,\dots,m_H}, \nonumber\\
&& U(m_1,\dots,m_H)=f(m_1,\dots,m_H)-1 \nonumber\\ &&+\sum_{1\le
n\le H}
  \frac{L_n}{L}\sqrt{1-m_n^2}.
\end{eqnarray}
{In Eq.(9) we take the maximum in the domain $-1\le m_n\le 1$. The
function $U(m_1,..m_H)$ is the equivalent of the potential in
classic mechanics.}

Following \cite{sa07}, we identify the mean fitness (the maximum
eigenvalue of the matrix on the right hand side of Eq.(\ref{e4}))
with the lower bound of Eq.(\ref{e9}),
\begin{eqnarray}
\label{e10} R= max[U(m_1,\dots,m_H)]|_{m_1,\dots,m_H}.
\end{eqnarray}
One can calculate the mean fitness $R$ by differentiating the
function $U(m_1,\dots,m_H)$.

Thus, we defined the mean fitness for the general multi-dimensional
mean-field like fitness landscape for the evolution model with
selection and mutation.

Figure 1 gives the comparison of our analytical result for
Eq.(\ref{e9}) with numerics of the 3-dimensional model.

\begin{figure}
\centerline{\includegraphics[width=0.95\columnwidth]{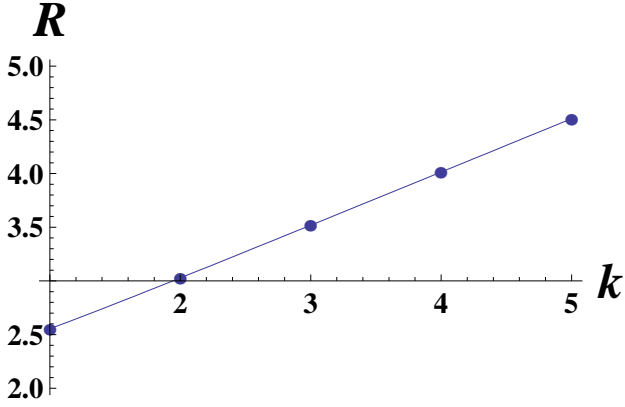}}
\caption{ The comparison of analytical result (smooth line) with the
numerics (dots) for the 3-d model with $L_1=L_2=L_3=20$. The whole
genome mutation rate is $1$.  The first part of the genome  has a
fitness $f_1(m_1)=km_1^2/2$. In the second part all the mutations
are lethal. For the fitness contribution from this part, we have a
zero for the sequences with zero mutations in this block and
$-\infty$ for the non-zero mutations.
 Part
three is described by a single peak fitness landscape with fitness
$J=3$ for the peak subsequence and zero for other subsequences.
Thus, the fitness function is defined as
$f(m_1,m_2,m_3)=km_1^2/2-[1-\delta(m_2-1)]*\infty+J\delta(m3-1)$,
where the discrete $\delta(x)$ function is equal to 1 at zero and is
equal to 0 otherwise.
 The mean fitness is given as $k(1-1/(3k))^2/2+3-2/3$. }
\label{fig1}
\end{figure}

\subsection{The Multidimensional Model with Lethal Mutations}

Let us now consider a model where there exists some probabilities of
lethal mutations: the Malthusian fitness $r$ (after $t$ period of
time, the population without mutation grows $e^{rt}$ times) while in
the parallel (Crow-Kimura) model is becoming -$\infty$ \cite{sa10b}.

 At any piece of the genome, we consider the master sub-sequence, having non-lethal
$L_n(1-\lambda)$ neighbors with single mutations, where $0\le
\lambda<1$ is a parameter describing the fraction of lethal
mutations. When the fitness is a function of the Hamming distance
from the reference sequence, we simplify the evolution equations
using this symmetry. We define some mutations from the reference
sequence as lethal mutations and
 assume
that any sequence having at least one lethal mutation (plus some
non lethal mutations) has a -$\infty $ fitness. Therefore at the
$l$-th Hamming class we have
$$N_{l,\lambda_n}=\frac{L_n(1-\lambda_n)!}{(L_n(1-\lambda_n)-l)!l!}$$
viable $l$ point mutants, and as a maximal $l$, we take
$L_n(1-\lambda_n)$. For a small $l\ll L_n$, there is a dilution of
the sequence space via a factor $(1-\lambda_n)$, while the total
number of viable sequence is:
\begin{equation}
\label{e11}
\sum_{l=0}^{L_n(1-\lambda_n)}N_{l,\lambda_n}=2^{(1-\lambda_n)L_n}.
\end{equation}

We define now the fitness function as
\begin{eqnarray}
\label{e12} r_{l_1,\dots,l_H}\equiv L f(m_1,\dots,m_H),
\end{eqnarray}
where instead of Eq.(\ref{e1}), we now define
\begin{eqnarray}
\label{e13} l_n=L_n\frac{1+m_n}{2}(1-\lambda_n).
\end{eqnarray}

Then the calculation is, identical to those in \cite{sa10b}, give
\begin{eqnarray}
\label{e14} R&=&max_m[f(m_1,\dots,m_H)-
1\nonumber\\
&+&\sum_n\frac{L_n}{L}(1-\lambda_n) \sqrt{1-m_n^2}].
\end{eqnarray}

\subsection{The Model in Multipeak Fitness Landscape}

We formulated the model by Eq.(3) for a rather general case. The
multi-peak model, considered in \cite{sa06a}, could be derived as
a particular case of our solution.

Let us choose $H=2^{K-1}$ and consider $K$ reference sequences
with our $s^n_i$ spins, $1\le i\le L,~1\le n\le H$. At any
position $i$ along the genome, we are looking at the alignment of
spins in our $K$ reference sequences. We have chosen the first
configuration with all $"+"$ spins and define the alignment of
spin along the $i$-th reference sequence at the $n$-th piece of
genome as $\alpha_{i,n}$. We  group together the configurations
$s^n_i=\alpha_{i,n}$ and $s^n_i=-\alpha_{i,n}$, where
$\alpha_{i,n}=\pm 1$ and these two cases have a joint probability
$ L_n/L$. The magnetization of the $i$-th sequence $M_i$ is
defined through our $m_n$ as:
\begin{eqnarray}
\label{e15} M_i=\sum_{n=1}^H\frac{L_n}{L}\alpha_{ni}m_n.
\end{eqnarray}
We then take a fitness which is a function of our $H$ reference
sequences. Thus, we should find the maximum of
\begin{eqnarray}
\label{e16}&& F(M_1,\dots,M_K)-1+\sum_{1\le n\le H}
  \frac{L_n}{L}\sqrt{1-m_n^2}\nonumber\\
  &&+\sum_ih_i[-M_i+\sum_{n=1}^H \frac{L_n}{L}\alpha_{ni}m_n]
\end{eqnarray}
where we introduced the auxiliary variables $h_i$. The maximum
condition gives
\begin{eqnarray}
\label{e17}&& h_i=\frac{\partial F(M_1,\dots,M_K)}{\partial
M_i},\nonumber\\
&& \sum_ih_i\alpha_{ni}=\frac{m_n}{\sqrt{1-m_n^2}}.
\end{eqnarray}
The last system of equations coincides with the one derived in
\cite{sa06a} with the mapping:
\begin{eqnarray}
\label{e18} m_n=\frac{1}{1+(\sum_{i=1}^K\alpha_{n,i}H_i)}
\end{eqnarray}
where $H_i$ are the fields, conjugate to the $M_i$ in
Eq.(\ref{e10}) of \cite{sa06a}.
 A single difference: in \cite{sa06a}
we defined ${L_n}/{L}$ for $2^{K}$ situations (misprints in
Eqs.(\ref{e9}) and (\ref{e24}) of \cite{sa06a}, in which $2^{K}$
should be replaced by $2^{K-1}$  ), instead of $2^{K-1}$ in the
current article.

\subsection{The 2-dimensional case}

{\bf The definition of the model }. Let us consider the 2
dimensional case. We have a system of equations:
\begin{eqnarray}
\label{e19} \frac{dp(l_1,l_2,t)}{dt}=
(r_{l_1,l_2}-L-LR)p(l_1,l_2,t)\nonumber\\
+\sum_{\beta=\pm 1}
L_1x_{\beta}(l_1,L_1)p(l_1-\beta,l_2,t)+\nonumber\\
L_2x_{\beta}(l_2,L_2)p(l_1,l_2-\beta,t),\nonumber\\
  R=\frac{1}{L}\sum_{0\le l_n\le L_n}p(l_1,l_2,t)r_{l_1,l_2},
\end{eqnarray}
We have a HJE for this case:
\begin{eqnarray}
\label{e20} \frac{\partial u(m_1,m_2)}{\partial t}=- H(m_1,m_2;
\frac{\partial u}{\partial_{m_1}},\frac{\partial
u}{\partial_{m_2}}),
\nonumber\\
-H(m_1,m_2;\hat P_1,\hat P_2)=f(m_1,m_2)
 \nonumber\\
-1+\sum_{1\le n\le 2} \frac{L_n}{L}(\frac{1+m_n}{2}e^{2\hat P_n}+
   \frac{1-m_n}{2}e^{-2\hat P_n})
\end{eqnarray}
The mean fitness R is defined through the equations
\begin{eqnarray}
\label{e21} R=f(m_1,m_2) -1+\sum_{1\le n\le 2}
\frac{L_n}{L}\sqrt{1-m_n^2},\nonumber\\
f'_1(m_1,m_2)=\frac{L_1}{L}\frac{m_1}{\sqrt{1-m_n^2}},\nonumber\\
f'_2(m_1,m_2)=\frac{L_2}{L}\frac{m_2}{\sqrt{1-m_n^2}}
\end{eqnarray}
{\bf The two-block model with lethal mutations}. In Fig. 2 we
compare the analytical results with the numerics for the two-block
model,where one part has the length $(L-n)$ with a lethal mutation
(all the spin configurations of the block besides the one have
$-\infty$ fitness), and the other block has the length n and a
fitness $f(m_1)=km_1^2/2$. We obtain the mean fitness of this model
as
\begin{eqnarray}
\label{e22} R=\frac{k}{2}(1-\frac{n}{k(n+m)})^2-\frac{m}{n+m}
\end{eqnarray}

\begin{figure}
\centerline{\includegraphics[width=0.95\columnwidth]{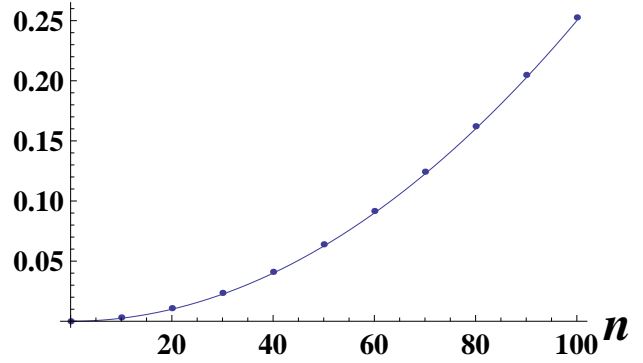}}
\caption{ The mean fitness R versus the length of the first block in
the 2-d model with a fitness $f(m_1)=m_1^2$ for the first block with
the length n and lethal mutations for the second block, with zero
fitness for the peak configuration of the second block. The total
length of the genome is $100$. The analytical results are given by
the smooth line.} \label{fig3b}
\end{figure}

{\bf The asymmetric original distribution}.
 We consider the
original distribution $m(0)=0.6$ for the symmetric distribution,
only considering the 1-d (Crow-Kimura) model:
\begin{eqnarray}
\label{e23} f(m)=\frac{k}{2}m^2
\end{eqnarray}
 Later we take the simplest asymmetric distribution, where the part
$L_1$ spins have $l_1$ minus spins and  original narrow distribution
with $m_1=1-2l_1/L_1$. Another part has $l_2$ minus spins have
original narrow distribution around $m_2=1-2l_2/L_2$. We consider
the model by Eq.(19) with the fitness
\begin{eqnarray} \label{e24}
f(m_1,m_2)=\frac{k}{2}m^2,\nonumber\\
m=(m_1\frac{L_1}{L}+m_2\frac{L_2}{L})
\end{eqnarray}
Fig.3 gives the results of the dynamics for $m,m_1,m_2$.

\begin{figure}
\centerline{\includegraphics[width=0.95\columnwidth]{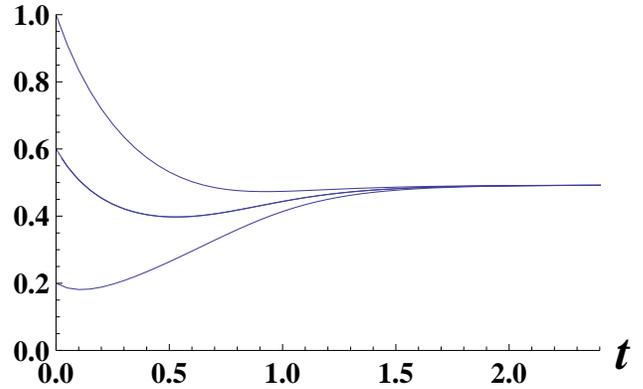}}
\caption{ The dynamics of $m,m_1,m_2$ for the model by Eqs.(19),(24)
with $m_1(0)=1,m_2(0)=0.2,L_1=L/2,L_2=L/2$. The middle line
corresponds to the m by Eq.(24) or the Crow-Kimura model by Eq.(23)
with $m(0)=0.6.$ } \label{fig4}
\end{figure}

{\bf The population distribution for the 2-d case}. Let us
investigate the population distribution. We consider a fitness
\begin{eqnarray}
\label{e25} f(m_1,m_2)= \frac{1}{2}\sum_{ij}A_{ij}m_im_j,\nonumber\\
A_{11}=k_1,A_{22}=k_2,A_{12}=k_3
\end{eqnarray}
Assuming an ansatz
\begin{eqnarray}
\label{e26}P(x)= \frac{\pi
L\sqrt{det(G)}}{2}\exp[-L\frac{<x|G|x>}{2}]\nonumber\\
u(x)=-\frac{<x|G|x>}{2}, \vec x=\vec m-\vec s
\end{eqnarray}
we obtain for the correlation:
\begin{eqnarray}
\label{e27}K_{ij}\equiv \int dx
\frac{p'_i(x)p'_j(x)}{p(x)}=\sum_{l,n}G_{li}G_{nj}<x_lx_n>=G_{ij}
\end{eqnarray}
Differentiating the HJE Eq.(20) for the steady state by $x_1,x_2$,
and putting $p_1=0,p_2=0$, we obtain:
\begin{eqnarray}
\label{e28} A_{ij}s_j-G_{ij}s_j
\end{eqnarray}
For the symmetric fitness case
\begin{eqnarray}
\label{e29} A_{11}=A_{22}=k_1,A_{12}=k_3\nonumber\\
G_{11}=G_{22}=g_1,G_{12}=g_3,
\end{eqnarray}
$s_1=s_2$ and Eq.(28) gives:
\begin{eqnarray}
\label{e30} k_1+k_3=g_1+g_3
\end{eqnarray}
We verified the validity of  Eq.(30) by the numerics in Table 1.

\begin{center}
\begin{table}[t]
\caption{Mean fitness for the 2-d model by Eq. (25). $L_1=L_2=L/2,k_2=k_1$ }%
\begin{tabular}{|c|c|c|c|c|c|c|c|}
\hline L                                 & 100   & 100   & 100 & 100     & 100  \\
\hline $k_1$                             & 4     & 8     & 4     &
4& 3
\\ \hline $K_3$                          & 3     & 2     & 5     & 6     & 7  \\
\hline $R_{theor}$                       &7.0312 &9.025  &8.0277 & 9.025 & 9.025   \\
\hline $R_{num}$                         &7.0315 &9.0251 &8.0280 &9.025 & 9.0251   \\
\hline $\frac{g_1+g_3}{k_1+k_3}_{theor}$ & 1.    & 1     & 1     & 1     & 1       \\
\hline $\frac{g_1+g_3}{k_1+k_3}_{num}$   & 0.991 & 0.984 & 0.992 & 0.993 & 0.994   \\

\hline
\end{tabular}%
\end{table}
\end{center}

\section{Recombination in a multi-dimensional fitness landscape}
\subsection{The Model}

In order to describe the recombination (horizontal gene transfer),
we follow  \cite{le05, de07}.
 We consider the
following system of equations:
\begin{eqnarray}
\label{e31} &&\frac{dp(l_1,\dots,l_H,t)}{dt}=
(r_{l_1,\dots,l_H}-LR)p(l_1,\dots,l_H)\nonumber\\
&& -Lp(l_1,\dots,l_H,t)
 \nonumber\\
&&+\sum_{\beta=\pm 1,~n}L_n
    x_{\beta}(l_n-\beta,L_n)p(l_1,\dots,l_n-\beta,\dots,l_H,t)\nonumber\\
   && +c[(\sum_{\beta=\pm 1,n}L_n
    x_{\beta}(l_n,L_n)\frac{1-\beta s_n}{2}-1)p(l_1,\dots,l_H,t)\nonumber\\
&& +\sum_{\beta=\pm 1, n}
   L_nx_{\beta}(l_n-\beta,L_n)\frac{1+\beta s_n}{2}\nonumber\\
  &&\times  p(l_1,\dots,l_n+\beta,\dots,l_H,t)],
\end{eqnarray}
where the sum over $n$ extends from 1 to $H$, and
\begin{eqnarray}
\label{e32}
s_{n}=\sum_{l_1...l_H}p(l_1,\dots,l_H,t)\frac{L_n-2l_n}{L_n}
\end{eqnarray}
is the equivalent of surplus or "surface" magnetization. For the
simple symmetric fitness landscape ($K=1$) which has one surplus
parameter, but now there are $H$ parameters.

 The term $-Lp(l_l...l_K,t)$ describes the
mutations of the whole genome with a rate $1$ per allele;
 the following line  describes the mutation. Using a coefficient $c$, we define the diagonal
 recombination terms: $-c$ is the total rate of changing
 the given  sequence, and $ x_{\beta}(l_n,L_n)\frac{1-\beta s_{n}}{2}$ describes the
 recombination event when we replace a spin from our current sequence
  with the same kind of spin from the pool of spins
 at the same position in population.
 In the second term inside ``[\dots]'', we define the recombination terms as the change in the
 current configuration: we replace a spin with an
 opposite spin from the spin pool.

Let us derive the Hamilton-Jacobi equation. We used the same anzats,
Eq. (\ref{e5}),  as before; the simple derivations give:
\begin{eqnarray}
\label{e33}
&&\frac{\partial u}{\partial t}=H(m_1,\dots,m_K;s_1,\dots,s_H;u'_1,\dots,u'_H), \nonumber\\
&&-H=f(m_1,\dots,m_H)-f(s_1,\dots,s_H)-1-c
 \nonumber\\
&&+\sum_{\beta=\pm 1,1\le n\le H}
    \frac{l_n}{L}(\frac{1+m_n}{2}e^{2u'_n}+\frac{1-m_n}{2}e^{-2u'_n})\nonumber\\
    &&+c(\sum_{n}\frac{l_n}{L}(\frac{(1+m_n)(1+s_n)}{4}+\frac{(1-m_n)(1-s_n)}{4})-1
    \nonumber\\
&&+\sum_{ n}\frac{l_n}{L}[\frac{(1+m_n)(1-s_n)}{4}e^{2u'_n}
+\frac{(1-m_n)(1+s_n)}{4}e^{-2u'_n}].\nonumber\\
\end{eqnarray}
where we denote $u_n=\frac{\partial u(m_1,\dots,m_H,t)}{\partial
m_n}$. The function $u(m_1,\dots,m_H,t)$ has the maximum at the
point $(m_1,\dots,m_H)=(s_1,\dots,s_H)$.

We don't see a simple way to calculate the asymptotic solution of
the last equation.

\subsection{An Approximate Solution of Recombination Dynamics}

 Let us  consider
the dynamics of the initial normal distribution,
\begin{eqnarray}
\label{e34}&&P(m_1,\dots,m_H,0)\nonumber\\
&&=\exp[-L\sum_{ln}\frac{y_{ln}}{2}(m_l-s_l(0))(m_n-s_n(0))].
 \end{eqnarray}
Equation (\ref{e34}) describes a narrow distribution around some
Hamming classes.

We assume that for some not too large periods of time, we have a
similar solution,
\begin{eqnarray}
\label{e35}&&P(m_1,\dots,m_H,t)\nonumber\\
&&=\exp[-L \sum_{ln}\frac{y_{ln}}{2}(m_l-s_l(t))(m_n-s_n(t))],
 \end{eqnarray}
where $y_{ln}$ describes the normal distribution.

 We get the following system of equations for $ds_n(t)/dt$
 using our Hamiltonian form Eq.(\ref{e33})
\begin{eqnarray}
\label{e36}
&&-\sum_ny_{ln}\frac{ds_n}{dt}=-\frac{dH(s_1,\dots,s_H;s_1,\dots,s_H;0,\dots,0)}{dm_l}\nonumber\\
&&+\sum_n\frac{dH(s_1,\dots,s_H;s_1,\dots,s_H;0,\dots,0)}{dp_n}y_{ln}.
 \end{eqnarray}
Let us prove that the last two terms do not depend on c. From  the
first line we obtain:
\begin{eqnarray}
\label{e37}
&&-\frac{dH(s_1,\dots,s_H;s_1,\dots,s_H;0,\dots,0)}{dm_l}\nonumber\\
&&=\frac{\partial f(m_1,\dots,m_H)}{\partial m_l}.
 \end{eqnarray}
For the rest we derive:
\begin{eqnarray}
\label{e38} -2\sum_l\frac{L_l}{L}s_ly_{ln}.
 \end{eqnarray}
Eventually, putting the results of Eqs.(\ref{e37}),(\ref{e38}) into
Eq.(\ref{e36}), we derive:
\begin{eqnarray}
\label{e39}\sum_ny_{ln}\frac{ds_n}{dt}=f'_n(s_1,\dots,s_H)-2\sum_n\frac{L_l}{L}s_ly_{ln}.
 \end{eqnarray}
 Thus for the initially narrow distribution of population by Eq. (\ref{e34})
 and mean-field like fitness landscape, the
recombination does not have any impact on the relaxation dynamics
for some period of time T. If the number of mutations and
recombination per genome per replication is in the order of 1, then
we have the following condition for this time period: $1\ll T\ll L$.

\subsection{Asymmetric Recombination.}

The theorem from the previous section is not valid for the
asymmetric recombination, since  we have different recombination
 rates for the allele changes to up and down.
 Consider the simple case of a one dimensional fitness landscape.
\begin{eqnarray}
\label{e40} &&\frac{dP_{l}}{dt}=\left[(r_l-LR) \right]P_{l}+
(l+1)P_{l+1}+(1-(l-1))P_{l-1}\nonumber\\
 &-& L
[c_1(1-\frac{\bar{l}}{L}) \frac{l}{L}P_{l} +
c_2\frac{\bar{l}}{L}(1-\frac{l}{L})P_{l}]\nonumber\\
&+&L[c_1(1-\frac{\bar{l}}{L})\frac{l+1}{L}P_{l+1}+
c_2\frac{\bar{l}}{L}(1-\frac{l-1}{L})P_{l-1}],
\end{eqnarray}
where $c_1,c_2$ describe the recombination rates to the up and down
directions in Hamming classes and $\bar l=\sum_lP_ll$.

Using an ansatz $P_l=\exp[Lu(m,t)]$, we derive the following HJE
\begin{eqnarray}
\label{e41} \frac{du}{dt}&=&
f(m)-f(s)-c_1\frac{(1+m)(1-s)}{4}-c_2\frac{(1-m)(1+s)}{4}\nonumber\\
&+&e^{2u'} \frac{1+m}{2}[1+\frac{1-s}{2}
c_2]-1\nonumber\\
&+&e^{-2u'}\frac{1-m}{2}[1 + \frac{1+s}{2} c_1].
\end{eqnarray}
Now we take $u(t)=-y(m-s(t))^2/2$ and get an equation
\begin{eqnarray}
\label{e42} y\frac{ds}{dt}=
f'(s)-2ys(t)-\frac{(c_1-c_2)}{2}(1-s(t)^2)y.
\end{eqnarray}
We see that the recombination immediately starts to change the
distribution, see Fig.4 for the illustration.

\begin{figure}
\centerline{\includegraphics[width=0.95\columnwidth]{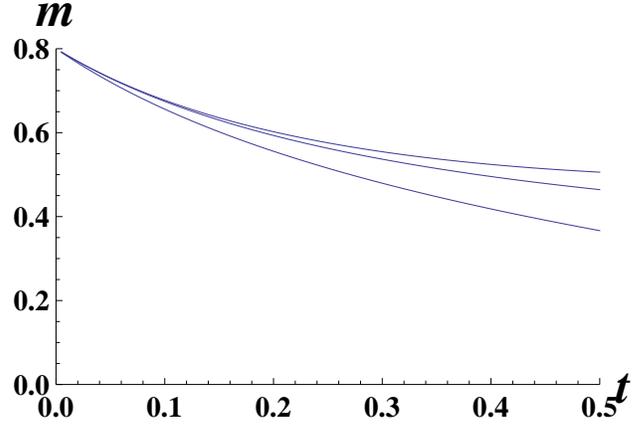}}
\caption{ The dynamics of $m\equiv 1-2n/L$, where $n$ is the mean
number of mutations in the model by Eqs. (40) and (41) with
$f(m)=m^2,L=1000$. The upper corresponds to the model without
recombination, the middle line to the model with symmetric
recombination with the rate $c=1$ and the low line to the
asymmetric recombination with $c_1=1.5,c_2=0.5$. The time scale is
chosen as in Eq.(41). For the zero selection case at time period 1
almost all the alleles in the genome are mutated.} \label{fig4b}
\end{figure}

\subsection{The Recombination Model with Lethal Mutations}
In order to  describe the lethal mutations, we consider the genome
which consists of two parts with the length $L_1=\lambda L$ and
$L(1-\lambda)$. In the first piece, there is only one sequence with
the fitness 0, and any mutation in this part gives a lethal sequence
with the $-\infty$ fitness.

We can investigate the situation using our model by Eq.(40).
Previously we used the mutation rate  $1$. Now we introduce the
mutation rate $\mu_0$ per nucleotide and $c$ as a recombination rate
per nucleotide.

We just write the equations for $p(0,l)\equiv p_l$, identifying
also $r(0,l)\equiv r_l$:
\begin{eqnarray}
\label{e43}&& \frac{dp_l}{dt}=r_lp_l
 -p_l\mu_0 L
 \nonumber\\
&&+ \bar L[\mu_0(\frac{l-1}{\bar L}p_{l+1}+\frac{\bar L-l+1}{\bar L}p_{l-1})\nonumber\\
    &&+ c(\frac{l-1}{\bar L}\frac{1+s_n}{2}+\frac{\bar L-l+1}{\bar L}\frac{1-s_n}{2}-1)p_l\nonumber\\
&&+ c(\frac{l-1}{\bar L}\frac{1-s_n}{2}p_{l-1}+\frac{\bar
L-l+1}{\bar L}\frac{1+s_n}{2}p_{l+1}),
\end{eqnarray}
where we denoted the length of the genome without lethal mutations
as $\bar L=L(1-\lambda)$.  While in the previous models we took
$\mu_0=1$, now we write formulas for general $\mu_0$.

 Let us  define
\begin{eqnarray}
\label{e44} m=\frac{2l-\bar L}{\bar L},\nonumber\\
r_l=f(m)\hat L
\end{eqnarray}
then we can use the results of \cite{de07} to calculate the mean
fitness. If we define the potential $U(m,s)$:
\begin{eqnarray}
\label{e45}
 &&U(m,s)=f(m)+\sqrt{(1-m^2)C}+\frac{cms}{2}-\frac{c}{2},\nonumber\\
 &&C=[(\mu_0+\frac{c}{2})^2-\frac{c^2s^2}{4}]
\end{eqnarray}
then the mean fitness of the genome is defined as
\begin{eqnarray}
\label{e46}
&& max[\bar L(U(m,s)-L\mu_0)],\nonumber\\
 &&LR=Lf(s)(1-\lambda).
\end{eqnarray}

\subsection{The finite population version of the model with lethal mutations.}

In the case of HIV, there are highly variable parts of the genome
with about $100$ nucleotides  \cite{bo05}. In \cite{bo05}  the use
of an evolution model with shorter effective genome length to
describe the virus evolution in such a case has been suggested;
later this idea was applied in \cite{pe08}. We assume that the usage
of an effective genome length is reasonable for the zero epistasis
case, while in the case of lethal mutations as well, we can not use
an ordinary model with the short genome length.

Extending the ideas in \cite{2012EPL}, we suggest the following
finite population versions of the model. The genome consists of two
parts. The first part has a length $L\lambda$ where all the
mutations are lethal, while the $n$ mutations from the part with the
length $L(1-\lambda)$ give a mutant with the fitness function $r_n$.
The population is described via $L-\bar n$ viable sequences and the
$\bar n$ lethal ones, and the total population size $N$ is fixed. We
describe the population via the number of viruses $n_l$ in the
$l$-th Hamming class, $0\le l\le L$ and $\bar n$. We have a
conserved population size, $n+\sum_{l=0}^Ln_l=N$.

During the time period $\delta t$ ,there are $\mu\delta
t(1-\lambda)$ non-lethal
 mutations and $\mu\delta t \lambda$ lethal mutations.

We consider
 the following steps during the evolution:

 a.  A birth of $\delta \bar n$  new
 lethal mutants which is a binomial random process with the  probability parameter $\delta t\lambda$ and  $(N-\bar
 n)$ trials.

 b. A birth of $\delta n_l$ new viruses in the l-th class, which is a
 binomial random process with $n_l$ trials and a probability parameter $r_l \delta
 t$.

 c. Forward non-lethal mutations $f_l$, which are described via binomial
 random process with a probability parameter $\delta
 t(1-\lambda)\frac{l}{L}$ and $n_l$ trials.

  Backward non-lethal mutations $b_l$, which described via the binomial random
  process with probability parameter $\delta
 t(1-\lambda)\frac{L-l}{L}$ and $n_l$ trials.

  Thus after these mutation processes, $n_l\to n_l-f_l-b_l$ ,
  $n_{l+1}=n_{l+1}+f_l$, $n_{l-1}=n_{l-1}+b_l$.

d. The dilution of the model, where we reduce the virus population
via $\bar n+\sum_{l=0}^L\delta n_l$ numbers, uniformly distributed
via $L+2$ classes.

 \section{ Conclusion}

We formulated and solved the evolution model on the multidimensional
fitness space, where we considered the genome as a collection of
several pieces and the total fitness as the function of the allele
type fractions of the pieces. Such a model is more general and more
realistic than the multi-point fitness landscape, considered in
\cite{sa06a}. The numerics confirmed our analytical results well.

 We calculated
  the mean fitness of this model, including the case of lethal
  mutations and found a simple way of deriving the results of the multi-peak fitness
  models.

We formulated the recombination model in the multidimensional
fitness space. While we could not calculate the mean fitness, we
derived the Hamiltonian-Jacobi equation for the dynamics of the
population and deduced an important theorem about the dynamics. For
the initially narrow initial distribution and mean-field fitness
landscape, the recombination does not affect the dynamics of the
population for a rather long period of times ( see Fig.2). This
theorem is not valid in the case of asymmetric recombination.

 We formulated  the finite population version of the model with
lethal mutations. Our results could be applied to model virus
experiments, prescribing to different parts of the genome either
lethal mutations or negative or positive selection. For example, we
can apply our model in the case of the Dengue virus, where 95\% of
the genome is epistasis free while there are strong correlations
between the gene contributions of the reminder 5 \% \cite{gr12}.

{The main open mathematical problem in the investigation of
multi-dimensional evolution is the calculation of the surplus and
the distribution around the peak of distribution. While we
calculated the mean fitness, we failed to calculate the surplus. In
classical mechanics, one can easily define the ground state energy
and the position of the interacting particles, looking for the
minimum of potential energy. Now, for our Hamiltonian by Eqs.(20),
the situation is highly non-trivial. One should consider the
asymptotic solution for the characteristics (the solutions of
Hamilton equation), looking for the steady states. Another
problem,which is important for applications,is to define the
quadratic expansion of the solution $u(m_1,m_2)$ near the maximum of
distribution. Again, the situation is highly non-trivial, and
different statistical physics phases are possible like the phases in
\cite{sa09}. While we found some relations, Eqs.(28),(30), we failed
to find the complete solution of distribution. We hope that it is
possible to succeed using the advanced methods of HJE, to address
this open problem.}

\section{acknowledgments}

Several years ago, C. Biebricher criticized our former choice of
fitness landscape \cite{sa06a}; he  pointed out to us the current
version of the fitness landscape as more adequate for virus
evolution. We thank R. H. Griffey  for the discussion of possible
application of our model to analyze  the evolution data of Dengue
virus. This work was supported by Academia Sinica, National Science
Council in Taiwan with Grant Number NSC 100-2112-M-001-003-MY2 and
National Center for Theoretical Sciences in Taiwan.

\end{document}